\documentclass{aa}  
\usepackage{graphicx}
\usepackage{txfonts}

\usepackage{xspace}
\usepackage{xcolor}
\newcommand{\SgrA}{Sgr~A$^\star$\xspace}
\newcommand{\Ka}{\,K$\alpha$\xspace}
\newcommand{\suzaku}{\textit{Suzaku}\xspace}
\newcommand{\chandra}{\textit{Chandra}\xspace}
\newcommand{\xmm}{\textit{XMM-Newton}\xspace}
\newcommand{\herschel}{\textit{Herschel}\xspace}
\newcommand{\asca}{\textit{ASCA}\xspace}
\newcommand{\snr}{G359.41$-$0.12\xspace}
\newcommand{\KS}{KS\,1741$-$293\xspace}
\newcommand{\ergs}{erg\,s$^{-1}$\xspace}
\newcommand{\Msol}{M$_\odot$\xspace}

\newcommand{\sline}{S\textsc{xv}\Ka}
\newcommand{\cm}{cm$^{-2}$\xspace}
\renewcommand{\deg}{$^\circ$}

\begin{document}

\title{Glimpses of the past activity of \SgrA\\inferred from X-ray echoes in Sgr~C}

\author{
  D.~Chuard\inst{1,2}\thanks{\email{dimitri.chuard@cea.fr}}
  \and R.~Terrier\inst{2}
  \and A.~Goldwurm\inst{1,2}
  \and M.~Clavel\inst{3}
  \and S.~Soldi\inst{2}
  \and \\M.~R. Morris\inst{4} 
  \and G.~Ponti\inst{5}
  \and M.~Walls\inst{6}
  \and M.~Chernyakova\inst{6,7}
}

\institute{
  Irfu/D\'epartement d'astrophysique, CEA Paris-Saclay, Orme des Merisiers, 91191 Gif-sur-Yvette, France
  \and APC, Univ. Paris Diderot, CNRS/IN2P3, CEA/Irfu, Obs. de Paris, USPC, 75205 Paris Cedex 13, France
  \and Space Sciences Laboratory, 7 Gauss Way, University of California, Berkeley, CA 94720-7420, USA
  \and Dep. of Physics and Astronomy, University of California, Los Angeles, CA 90095, USA
  \and Max-Planck-Institut f\"ur extraterrestrische Physik, 85748, Garching, Germany
  \and School of Physical Sciences, Dublin City University, Glasnevin, Dublin 9, Ireland
  \and Dublin Institute of Advanced Studies, 31 Fitzwilliam Place, Dublin 2, Ireland
}

\date{Received / Accepted}

\abstract
    {For a decade now, evidence has accumulated that giant molecular
      clouds located within the central molecular zone of our Galaxy
      reflect X-rays coming from past outbursts of the Galactic
      supermassive black hole. However, the number of illuminating
      events as well as their ages and durations are still unresolved
      questions.}
    {We aim to reconstruct parts of the history of the supermassive
      black hole \SgrA by studying this reflection phenomenon in the
      molecular complex Sgr~C and by determining the line-of-sight
      positions of its main bright substructures.}
    {Using observations made with the X-ray observatories \xmm and
      \chandra between 2000 and 2014, we investigated the variability
      of the reflected emission, which consists of a Fe\Ka line at
      6.4~keV and a Compton continuum. We carried out an imaging and a
      spectral analysis. We also used a Monte Carlo model of the
      reflected spectra to constrain the line-of-sight positions of
      the brightest clumps, and hence to assign an approximate date to
      the associated illuminating events.}
    {We show that the Fe\Ka emission from Sgr~C exhibits significant
      variability in both space and time, which confirms its
      reflection origin. The most likely illuminating source is
      \SgrA. On the one hand, we report two distinct variability
      timescales, as one clump undergoes a sudden rise and fall in
      about 2005, while two others vary smoothly throughout the whole
      2000$-$2014 period. On the other hand, by fitting the Monte
      Carlo model to the data, we are able to place tight constraints
      on the 3D positions of the clumps. These two independent
      approaches provide a consistent picture of the past activity of
      \SgrA, since the two slowly varying clumps are located on the
      same wavefront, while the third (rapidly varying) clump
      corresponds to a different wavefront, that is, to a different
      illuminating event.}
    {This work shows that \SgrA experienced at least two powerful
      outbursts in the past 300 years, and for the first time, we
      provide an estimation of their age. Extending this approach to
      other molecular complexes, such as Sgr~A, will allow this
      two-event scenario to be tested further.}

\maketitle


\section{Introduction}
From our terrestrial outlook, the astrophysical phenomena occurring at
the centre of the Milky Way are observable at an unrivalled level of
detail because the nucleus of our Galaxy is the closest we can study
by two orders of magnitude. Consequently, the centre of the Milky Way
has emerged as a key prototype for the study of galactic nuclei. In
particular, like most massive galaxies, the Milky Way hosts a
supermassive black hole at its centre, named after its electromagnetic
counterpart, the compact radio source Sagittarius~A$^\star$ (\SgrA)
discovered by \cite{balick1974}. It is located around 8~kpc from
Earth, and its mass is about four~million times that of the Sun
\citep[see][for recent estimates]{chatzopoulos2015,boehle2016}. Nevertheless, in contrast
with active galactic nuclei (AGN), its emission is extremely faint. In
X-rays (2--10~keV), its absorption-corrected quiescent luminosity is
about $10^{33}$~\ergs \citep{baganoff2003}, which is not only several
orders of magnitude below the Eddington limit, but also far lower than
what is expected from black-hole feeding by the stellar winds in its
vicinity \citep{genzel2010}. A strong effort is being made to account
for this underluminous state using numerical models of the gas
dynamics \citep[e.g.][]{cuadra2015,moscibrodzka2017}. In particular,
radiatively inefficient accretion flow (RIAF) models are in very good
agreement with the observations of the steady-state emission from
\SgrA \citep{wang2013}.

The black hole also undergoes regular flare-like events, especially in
X-rays \citep[][and references
  therein]{neilsen2013,ponti2015flares,yuan2016,ponti2017}. During
these episodes, its luminosity can increase by up to two orders of
magnitude above the quiescent value. This variability indicates that
significant changes in the accretion flow, such as stochastic
acceleration, shocks, magnetic reconnection, or tidal disruption of
small bodies, are possible. By extrapolation, we may therefore infer
that \SgrA sporadically ventures out of its current low-luminosity
state. The idea that this could have occurred in the past motivates
the search for relics of potential past high-activity episodes in the
interstellar medium surrounding the central black hole \citep[see][for
  a review]{ponti2013}.

When an X-ray source experiences an intense burst, it is indeed
possible to track it long after it ends by monitoring its reflection
on any optically thick molecular material located along the trajectory
of the photons. This reflected emission consists of a strong
fluorescent line of neutral and low-ionised iron (Fe\Ka) at 6.4~keV,
together with much weaker lines of lower atomic number elements and a
continuum component produced by Compton scattering. In the case of
potential past flares of \SgrA, this emission is expected to come
primarily from the giant molecular clouds populating the central
molecular zone \citep[CMZ;][]{morris1996,sunyaev1993}.

In 1994, the \asca mission first detected a strong Fe\Ka signal in
Sgr~B2, the most massive molecular cloud in the Galaxy, which is
located at a projected distance of about 100~pc from \SgrA
\citep{koyama1996}. Since then, similar detections have been reported
at many other locations in the CMZ
\citep[e.g.][]{nobukawa2008,terrier2017}. Notwithstanding that this
non-thermal emission can also be produced by low-energy cosmic rays
\citep[LECR; e.g.][]{yusefzadeh2002,dogiel2009,tatischeff2012}, the
great variability observed in the Sgr~A complex
\citep{muno2007,ponti2010,capelli2011,clavel2013,clavel2014} and in
Sgr~B2 \citep{koyama2008,inui2009,terrier2010,nobukawa2011} is a solid
argument for a reflection origin.

Considering the estimated energetics of the illumination, the most
plausible explanation is that the source is \SgrA. Therefore, it is
possible to probe the past activity of the Galactic supermassive black
hole over the past few centuries by monitoring the echoes of its past
flares while they propagate through the CMZ. The current distribution
and evolution of the 6.4~keV bright clumps indeed suggest that \SgrA
experienced at least one, and probably two, powerful outbursts
($L\sim10^{39}$~\ergs) in the past few centuries
\citep{clavel2013}. However, this does not allow a proper
reconstruction of the past light curve of \SgrA as long as the
distances between the bright clumps and the black hole, and hence the
propagation delay, remain unknown. Two main approaches have emerged to
address this problem: correlating variations between multiple regions
all over the CMZ while assuming that they are illuminated by the same
event on the one hand \citep{clavel2013,churazov2017,terrier2017}, and
using the physical properties of the X-ray emission to constrain the
line-of-sight positions of individual clumps on the other hand
\citep{capelli2012,ryu2013,walls2016}.

From this standpoint, the molecular complex Sgr~C is a highly valuable
object of study as it allows the two methods to be applied together
for the first time. Sgr~C is indeed a suitable candidate for
three-dimensional position determination based on spectral analysis as
its clumps are well resolved. It is also ideally located for studying
correlations in the Fe\Ka emission from both sides of the Galactic
plane, since Sgr~C and Sgr~B2 are on opposite sides of \SgrA at
similar projected distances.

Fe\Ka line emission at 6.4~keV was first detected in Sgr~C by
\cite{murakami2001} with \asca. It was then resolved into four main
clumps by \cite{nakajima2009} with \suzaku. Furthermore, the study of
the thermal diffuse X-ray emission by \cite{tsuru2009} revealed an
elliptical object designated as \snr and an adjacent chimney-like
structure. These two features, which are notably bright in the \sline
line at 2.45~keV, are thought to be a supernova remnant (SNR)
candidate and its associated outflow. Many non-thermal radio filaments
are also found in the area, including one of the brightest in the CMZ,
the Sgr~C filament \citep{liszt1985,anantharamaiah1991,larosa2000}, as
well as two non-thermal X-ray filaments that may be two pulsar wind
nebulae \citep{chuard2017}. Because of the possible interaction of all
these structures with the $6\times10^5$~\Msol of molecular gas
contained in Sgr~C \citep{liszt1995}, some of its Fe\Ka emission may
be due to cosmic-ray irradiation rather than X-ray
reflection. Therefore Sgr~C offers a unique opportunity to study these
two competing scenarios for the origin of the Fe\Ka emission.

We study the variability of the Fe\Ka emission in observations of
Sgr~C made with the X-ray observatories \xmm and \chandra between 2000
and 2014 (Section~\ref{observations}). To do so, we used both imaging
analysis and light curve extraction
(Section~\ref{specanalysis}). Based on our results, we discuss the
plausibility of the reflection scenario compared to the cosmic-ray
irradiation scenario. Finally, by comparing our data to Monte Carlo
simulated reflection spectra, we are able to place the best
constraints to date on the line-of-sight positions of the main bright
clumps of Sgr~C (Section~\ref{montecarlo}). Ultimately, extending this
approach with the inclusion of other molecular complexes allows us to
partially reconstruct the past light curve of the Galactic
supermassive black hole (Section~\ref{pastactivity}).

\section{Observations and data reduction}
\label{observations}
Sgr~C has been repeatedly observed with \xmm and \chandra, during
either dedicated pointings or CMZ scans. We consider here all the
available observations from these two satellites, including the latest
\chandra observation that we were granted in 2014. All data were taken
with focal plane imaging spectrometers using X-ray CCDs, namely the
European Photon Imaging Camera (EPIC) onboard \xmm
\citep{turner2001,struder2001} and the \chandra Advanced CCD Imaging
Spectrometer \citep[ACIS;][]{garmire2003}. They represent a final
dataset of 14 observations, covering the period from September 2000 to
August 2014 (Table~\ref{tab:obs}).

\begin{table*}
  \caption{List of all \xmm (EPIC) and \chandra (ACIS-I) observations,
    grouped into the six periods we defined to study the time
    variability (Fig.~\ref{fig:lightcurves}).}
  \label{tab:obs}
  \centering
  \begin{tabular}{c c c c c }
    \hline\hline
    Date & Obs. ID & \multicolumn{3}{c}{Exposure Time (ks)}\\
    &         & EPIC-MOS & EPIC-pn & ACIS-I \\
    \hline
    2000-09-11 & 0112970701 & 23.89 & 20.00 & -- \\
    2000-09-21 & 0112970801 & 23.89 & 20.00 & -- \\
    \hline
    2005-07-22 & 5892 & -- & -- & 97.91  \\
    \hline
    2006-02-27 & 0302883101 & 11.47 & 9.84 & -- \\
    2006-09-09 & 0302884501 & 8.42  & 6.79 & -- \\
    \hline
    2007-09-06 & 0504940701 & 6.67 & 5.06 & -- \\
    2008-03-04 & 0511001301 & 6.67 & 5.06 & -- \\
    2008-09-27 & 0511001401 & 6.67 & 5.03 & -- \\
    \hline
    2012-08-30 & 0694640201 & 46.67 & 45.04 & -- \\
    2012-09-07 & 0694640101 & 43.67 & 42.04 & -- \\
    2012-09-12 & 0694640901 & 44.67 & 43.04 & -- \\
    \hline
    2014-07-29 & 16174 & -- & -- & 30.10 \\
    2014-08-01 & 16642 & -- & -- & 29.81 \\
    2014-08-03 & 16643 & -- & -- & 35.62 \\
    \hline
  \end{tabular}
\end{table*}

The data reduction was carried out using the standard tools provided
for each observatory, the \xmm Extended Source Analysis Software
\citep[ESAS;][]{snowden2008} included in the \xmm Science Analysis
Software (SAS) version 12.0.1, and the \chandra Interactive Analysis
of Observations software \cite[CIAO;][]{fruscione2006} version 4.8,
respectively.

Exposure-corrected images were created from \chandra data using the
CIAO scripts \texttt{fluximage} and \texttt{merge\_obs}. Only ACIS-I
chips were considered. We excluded events within a circular region of
radius $2.76'$ centred on $(l,b)=(359.56$\deg$,-0.08$\deg$)$ in the
2005 dataset because of contamination from the X-ray binary \KS, which
was very active at that time \citep{degenaar2013}. Maps of the Fe\Ka
emission were generated by integrating the counts in the band
6.32$-$6.48~keV and continuum-subtracted following the approach of
\cite{ponti2010} and \cite{clavel2013}. The underlying continuum
emission was estimated from images created in the 4.0$-$6.1~keV band,
assuming a power-law spectrum of photon index $\Gamma=2$ (i.e. with a
rescaling factor of 0.045).

Source spectra were extracted from \chandra data using the CIAO
\texttt{specextract} routine, which also generated the ancillary
response file (ARF) and the redistribution matrix file (RMF). It was
also used to extract background spectra from blank sky event files,
which were created by combining observations of relatively empty
fields available from the \chandra calibration database (CALDB version
4.7.2). \xmm spectra were extracted with the ESAS \texttt{mos-spectra}
and \texttt{pn-spectra} scripts. Filter-wheel closed event lists from
the ESAS calibration database were used to estimate the quiescent
particle background. All spectra were then analysed with
\textit{Sherpa}, the modelling and fitting package of CIAO
\citep{freeman2001}. They were rebinned until the square root of the
number of counts in each bin exceeded a minimum signal-to-noise ratio,
between 3 and 10, depending on the quality of the spectrum. The
analysis was restricted to the energy range 2$-$7.5~keV. Model fits
were carried out using a chi-square statistic with the Gehrels
variance function \citep{gehrels1986}. In the following, all errors
are given at $1\sigma$ (68\% confidence) level and descriptions
implicitly refer to Galactic coordinates.

\section{Variability of the Fe\Ka emission}
\label{specanalysis}
\begin{figure*}
  \centering
  \includegraphics[width=0.95\hsize]{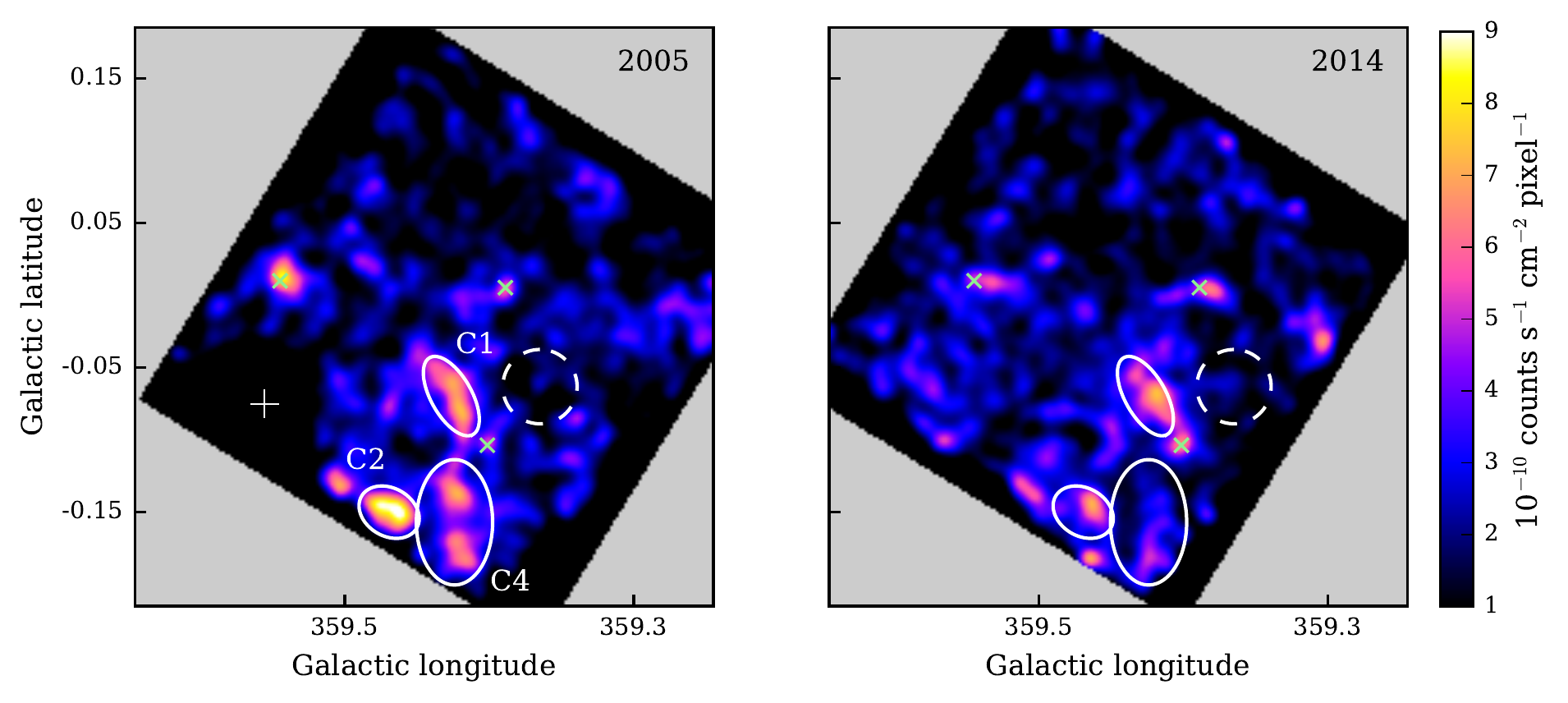}
  \caption{\chandra continuum-subtracted images of the Fe\Ka emission
    in Sgr~C for the observations taken in 2005 (\textit{left}) and
    2014 (\textit{right}). The maps are in units of
    counts\,s$^{-1}$\,cm$^{-2}$\,pixel$^{-1}$ with a pixel size of
    about $1''$, and smoothed using a Gaussian kernel of 20-pixel
    radius. The regions of interest (Sgr~C1, Sgr~C2, and Sgr~C4) are
    marked by the solid ellipses (see Table~\ref{tab:regions}). The
    dashed circle shows the control region. The white cross marks the
    position of the X-ray transient \KS, which has been removed from
    the 2005 dataset. The green crosses mark the positions of bright
    point sources
    ($F_{0.5-8\,\text{keV}}>10^{-5}$~counts\,s$^{-1}$\,cm$^{-2}$) from
    the catalogue of \cite{muno2006} that remain visible in the
    6.4~keV band after continuum subtraction.}
  \label{fig:img}
\end{figure*}

We first focus on the spatial distribution of the Fe\Ka emission in
Sgr~C. In order to identify the brightest regions and track them over
time, we took advantage of \chandra's unique imaging capabilities
thanks to its high angular resolution. We produced two images of the
Fe\Ka emission from our dataset from the data taken in 2005 and 2014,
respectively (Fig.~\ref{fig:img}). Both correspond to an exposure time
of about 100~ks (see Table~\ref{tab:obs}).

Three main bright regions are visible in the 2005 image. They match
those identified by \cite{nakajima2009} using \suzaku data taken in
Feburary 2006 (seven months later than the first ACIS-I \chandra
data). These regions (indicated as Sgr~C1, Sgr~C2, and Sgr~C4; see
also Table~\ref{tab:regions}) are known to be coincident with
molecular structures seen in radio
\citep{ryu2013,terrier2017}. Between 2005 and 2014, the morphology of
the Fe\Ka emission underwent noticeable changes. The image built from
the 2014 data reveals that there is almost no more emission in
Sgr~C4. Sgr~C2, the brightest region in 2005, is also clearly less
luminous, and the centroid of the bright area is shifted towards the
west. The case of C1 is more complicated as the peak value of the
radiance remains roughly constant, but the bright area shrinks and
moves towards the west as well. Thus, all three regions exhibit clear
variability. These results based on \chandra data are therefore in
full agreement with the findings of \cite{terrier2017} using \xmm
observations.

\begin{table}
  \caption{Galactic coordinates of the elliptical regions, following
    the naming convention of \cite{ryu2013} and
    \cite{terrier2017}. Sgr~C3 is out of the \chandra field of view
    and therefore not considered here.}
  \label{tab:regions}
  \centering
  \begin{tabular}{c c c c c}
    \hline\hline
    Name & $l$ (\deg) & $b$ (\deg) & Axes ($'$) & Angle (\deg) \\
    \hline
    Sgr~C1 & 359.43 & $-$0.07 & 3.66, 1.69 & 119 \\
    Sgr~C2 & 359.47 & $-$0.15 & 2.64, 1.99 & 150 \\
    Sgr~C4 & 359.42 & $-$0.16 & 5.20, 3.17 & 90 \\
    Control & 359.36 & $-$0.06 & 3.08, 3.08 & -- \\
    \hline
  \end{tabular}
\end{table}

In order to obtain more quantitative information about the
variability, we carried out a spectral analysis in the three main
regions identified above (Table~\ref{tab:regions}). Following the
standard approach, we fitted the spectra extracted in these regions
with a phenomenological model composed of a reflected emission
component and two thermal plasma components
\citep[APEC;][]{smith2001}. The reflected emission was modelled by an
absorbed (with fixed column density of $N_H=10^{23}$~\cm) power law of
photon index $\Gamma=2$ and a Gaussian line with $E=6.4$~keV and
$\sigma_E=10$~eV. The power law corresponds to the spectrum of the
illuminating source, and the absorption is the result of the crossing
of the molecular cloud. Following \cite{koyama2007} and
\cite{tsuru2009}, the temperatures of the two APEC models were fixed
at 1.0 and 6.5~keV, respectively, the first accounting for the soft
local plasma emission and the second for the hot Galactic ridge
emission \citep{worrall1982}. No fixed normalisation ratio was assumed
between the two APEC components, which were thus left free to vary. We
fixed the metallicity to solar values
\citep[following][]{nakajima2009} and applied foreground interstellar
absorption (with column density fixed at $N_H=7.5\times10^{22}$~\cm)
to all components except for the Gaussian line:
\begin{equation}
  \texttt{wabs}_{\texttt{1}}\times(\texttt{apec}_{\texttt{1}}+\texttt{apec}_{\texttt{2}}+\texttt{wabs}_{\texttt{2}}\times\texttt{powerlaw})+\texttt{gaussian}
\end{equation}
Even if this description is not fully physically relevant (see
Section~\ref{montecarlo}), it allows us to precisely measure the flux
in the Fe\Ka line.

The dataset including all available \xmm and \chandra observations was
split into six periods (see Table~\ref{tab:obs}) in order to study the
flux variability over time. For each region, the observations for all
these six periods were fitted simultaneously, leaving the
normalisation of the reflection components free to vary from one
period to another, while the thermal components were held constant
over time. The same approach was then applied to a control region
(Table~\ref{tab:regions}) that did not overlap the bright clumps in
order to estimate the level of background emission at 6.4~keV. The
reduced chi-square is very close to 1 for all fits.

\begin{figure*}
  \centering
  \includegraphics[width=0.95\hsize]{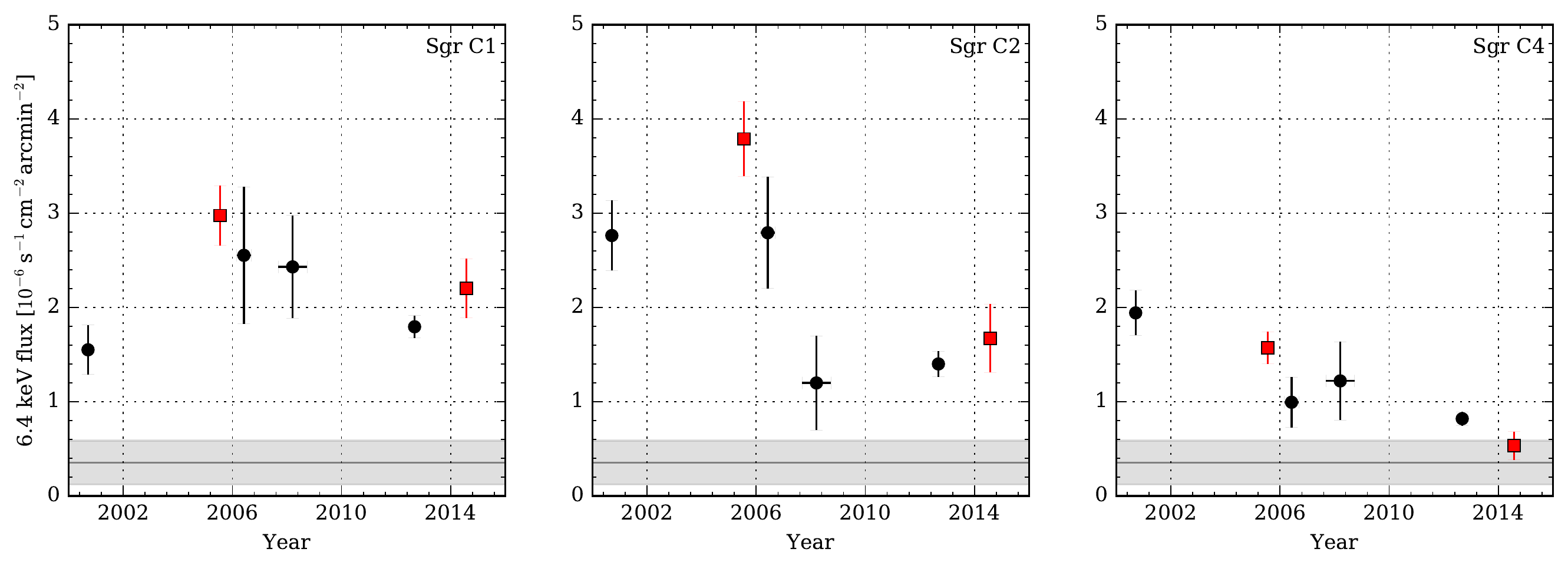}
  \caption{Light curves of the 6.4~keV line emission obtained by
    fitting the phenomenological model and integrated over the regions
    Sgr~C1 (\textit{left}), Sgr~C2 (\textit{centre}) and Sgr~C4
    (\textit{right}). The black circles and the red squares correspond
    to \xmm and \chandra observations, respectively. The grey shaded
    bars show the average level of background emission measured in the
    control region (their thickness represents one standard
    deviation).}
  \label{fig:lightcurves}
\end{figure*}

The Fe\Ka light curves we derived are shown in
Fig.~\ref{fig:lightcurves}. There is no evidence of a systematic shift
between \xmm and \chandra data points, suggesting that
intercalibration errors, although possible, do not significantly
affect our results. The observed trend confirms that the reflected
emission from Sgr~C has varied significantly for the past 15 years. In
Sgr~C2 and C4, the hypothesis that the flux has been constant during
the entire period is rejected at 5.6 and $5.4\sigma$,
respectively. The light curve of Sgr~C2 exhibits a maximum in 2005,
followed by a sharp decrease and a residual flux consistent with
constant emission from 2008 to 2014. However, since no data were
collected between 2000 and 2005, it is possible that the flux of
Sgr~C2 was even higher during this period. The light curve of Sgr~C4
is compatible with a linear decrease over the whole period down to the
background level indicated by the grey band. As for Sgr~C1, its light
curve would have been compatible with a constant without the excess
seen by \chandra in 2005 ($+62$\% compared to the 2005-excluded
average). In this respect, it is worth noting that the variability we
infer from the light curves only takes changes in intensity into
account and thus might be hiding changes in morphology within the
region of integration \citep{clavel2013}. This appears to be
especially true for Sgr~C1, whose 6.4~keV emission morphology evolves,
but with almost constant brightness.

The timescale and the amplitude of the variability detected at 6.4~keV
exclude the cosmic-ray irradiation scenario and confirm the reflection
origin of the Fe\Ka emission in Sgr~C, in agreement with
\cite{terrier2017}. The fluxes in the line and in the reflected
continuum appear to be positively correlated at the 4$\sigma$ level
\citep[determined by permutation following][to take uncertainties in
  both variables into account]{legendre1998}, which is also consistent
with the reflection scenario. Furthermore, the apparent motion of the
emission centroids towards the west, that is, away from \SgrA, may be
evidence of the signal propagation within these clouds. In the
following, we therefore consider that \SgrA is the source at the
origin of this reflected emission (see also Section~\ref{discussion}).

\section{Constraints on the past activity of \SgrA}
\subsection{Determining the line-of-sight positions of the clumps}
\label{montecarlo}
It is possible to use the observations of Sgr~C to place constraints
on the past light curve of \SgrA on the condition that we can
precisely determine the positions of the bright clumps along the line
of sight. This information can be extracted from the cloud X-ray
spectra but, until recently, no relevant physical models were
available for this purpose. Either phenomenological models (like the
one we used in Section~\ref{specanalysis}) or models developed for
other geometries, e.g. \texttt{MyTorus} \citep{zhang2015,mori2015} and
\texttt{pexrav} \citep{ponti2010}, were used. These poorly suited
models stongly limit any inference of the physical parameters of the
reflection phenomenon and may provide results that are marred by
significant systematic errors. Fortunately, Monte Carlo models
computing the spectrum produced by X-ray reflection from a spherical
molecular cloud have been recently developed. They are thus the
best-suited models available to date to physically describe the
reflection phenomenon.

We used a Monte Carlo spectral model developed by \cite{walls2016},
hereafter referred to as the Monte Carlo model, to determine the
line-of-sight positions of the reflecting clouds from their
spectra. This model has been designed to take the geometry of the
reflection, which has a major influence on the flux and spectral shape
of the cloud emission \citep{walls2016}, into account. The geometry is
parametrised in the model by the angle between the cloud, the
illuminating source, and the observer, referred to as the
line-of-sight angle. The total continuum flux at low energies notably
increases with this angle because photons only superficially penetrate
the cloud before being scattered towards the observer. Consequently,
the scattered photons are more likely to be absorbed in the low-angle
case than in the high-angle case. The line-of-sight angle, as well as
the cloud column density, also affect the strength of the Fe\Ka line
and the depth of the iron edge \citep{walls2016}. All these effects
are of key interest for the line-of-sight position determination.

\begin{figure}
  \centering
  \includegraphics[width=0.95\hsize]{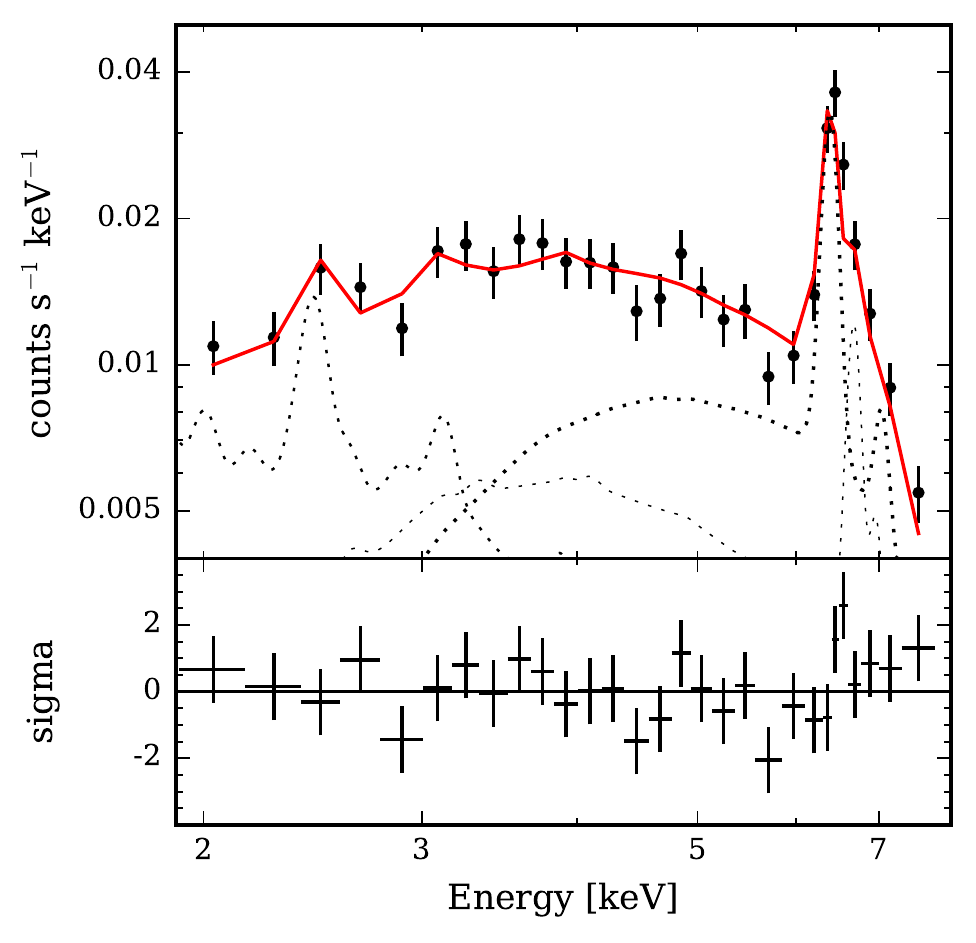}
  \caption{Spectrum of Sgr~C1 observed with EPIC-pn onboard \xmm in
    2012. The red line corresponds to the best-fit model. The dotted
    lines show the three components of the model: the hot
    (\textit{thin line}) and soft (\textit{medium}) thermal plasmas,
    and the Monte Carlo reflected emission (\textit{thick}).}
  \label{fig:spec}
\end{figure}

The line-of-sight angle and the cloud column density are free
parameters of the Monte Carlo model that can thus be constrained
through a spectral fitting procedure based on our dataset. In order to
have the best possible constraints on their values, we restricted our
analysis to the deepest observations (2000, 2005, 2012, and 2014). We
used the same model as in Section~\ref{specanalysis}, except that the
reflected emission was now modelled using the Monte Carlo model:
\begin{equation}
  \texttt{wabs}\times(\texttt{apec}_{\texttt{1}}+\texttt{apec}_{\texttt{2}}+\texttt{montecarlo})
\end{equation}
The parameters of the two thermal plasma components and of the
foreground interstellar absorption were kept unchanged (see
Section~\ref{specanalysis}). For a given region and a given period,
the normalisation was left free to vary. The line-of-sight angle and
the cloud column density (with a uniform density profile) were also
left free but constant over all periods. The dataset allows good
parameter constraints with satisfactory fit quality (see
Fig.~\ref{fig:spec} and Table~\ref{tab:fitres}). The variability of
the reflected component is found to be consistent with the light
curves obtained in Section~\ref{specanalysis}.

\begin{table}
  \caption{Values of the line-of-sight angles and uniform cloud column
    densities, obtained by fitting the \xmm and \chandra data with the
    Monte Carlo model. The metallicity is set to solar values.}
  \label{tab:fitres}
  \centering
  \begin{tabular}{c c c c}
    \hline\hline
    Region & Angle (\deg) & $N_H$ ($10^{23}$\cm) & $\chi^2$/d.o.f. \\
    \hline
    Sgr~C1 & $102.0_{-11.5}^{+5.8}$ & $2.18_{-0.37}^{+0.20}$ & $423.6/372$ \\
    Sgr~C2 & $66.7_{-6.3}^{+9.9}$   & $7.0_{-1.2}^{+1.2}$    & $234.0/206$ \\
    Sgr~C4 & $96.0_{-9.5}^{+6.2}$   & $1.63_{-0.13}^{+0.12}$ & $436.2/420$ \\
    \hline
  \end{tabular}
\end{table}

The best-fit values of the cloud parameters are given in
Table~\ref{tab:fitres}. The column densities we find are higher than
previous estimates \citep{yusefzadeh2007,nakajima2009}. As
\cite{walls2016} reported similar findings for Sgr~B2, this suggests
that prior modelling efforts might have been biased towards
underestimating column densities. In the case of Sgr~C2, the
difference is almost one order of magnitude. However, our values are
in the range of those inferred from the CS $J=1$--$0$ line emission
\citep[$N_H\sim 10^{23}$~\cm;][]{tsuboi1999} and from \herschel
far-infrared data \citep[$N_H\sim 10^{24}$~\cm;][]{molinari2011}.

\begin{figure}
  \centering
  \includegraphics[width=0.95\hsize]{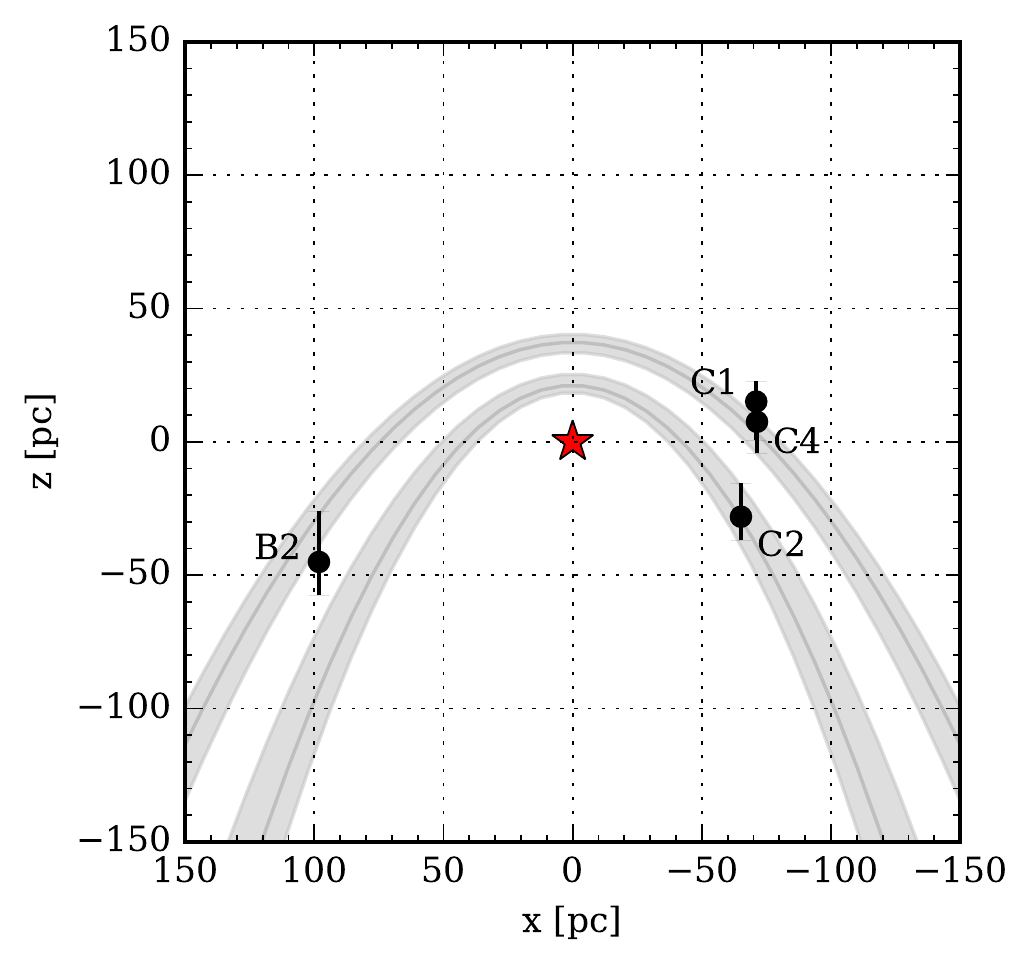}
  \caption{Face-on view of the Galactic centre. The negative direction
    along the $z$-axis points towards Earth. The red star marks the
    position of \SgrA. The black dots show the best-fit positions for
    the bright clumps Sgr~C1, C2, C4 (Table~\ref{tab:fitres}), and
    Sgr~B2 \citep{walls2016} in the solar-metallicity and uniform
    density profile case. The grey parabolas trace the best-fit
    associated wavefronts \citep[as seen from Earth;][]{sunyaev1998}
    for the two-event model. The width of the parabolas represents the
    statistical uncertainty on the position, not the duration of the
    associated event.}
  \label{fig:geom}
\end{figure}

Using the angles obtained from the fits and the celestial coordinates,
we are able to derive the 3D positions of the clumps within the CMZ
(Fig.~\ref{fig:geom}). Sgr~C2 is found to be in front of \SgrA , while
C1 and C4 are slightly behind it, both at comparable line-of-sight
distances. Although the error bars look rather small, one should keep
in mind that some systematics may affect our results, notably because
of the uncertainty on the cloud metallicity, its density profile, and
its geometry.

We tested the effect of the cloud metallicity $Z$. Three values that
are higher than solar are parametrised in the model ($1.3$, $1.7,$ and
$1.9$) but we only considered the $Z=1.3$ case since the Galactic
centre abudances are known to be close to solar \citep[see
  e.g.][]{davies2009}. We find than increasing the metallicity tends
to separate the clouds more from each other. In the case of Sgr~C1, we
observe that the line-of-sight angle increases with increasing $Z$ up
to $107.0_{-13.1}^{+9.2}$ degrees. Conversely, the angle of Sgr~C2
decreases to $60.0_{-8.2}^{+4.9}$ degrees. The position of Sgr~C4
remains almost unchanged ($93.0_{-15.1}^{+15.6}$ degrees). Although
some of these fits are statistically less good than those with $Z=1$,
this procedure allows us to estimate the typical systematic error that
results from the uncertainty on metallicity. On the whole, we find
that when we assume a higher metallicity, this does not fundamentally
alter the trend of our results even if it marginally changes the angle
values.

In addition, the shape of the cloud and its density profile might not
be as assumed by the Monte Carlo model. We tested Gaussian density
profiles. They have a very limited effect on our findings. The
line-of-sight angle changes by only a few degrees for Sgr~C1 and C2
and remains unchanged for Sgr~C4. Moreover, all the fits with Gaussian
profiles are statistically less good than those with uniform
density. Consequently, we did not consider these alternative profiles
further.

The cloud might also be only partially illuminated if the burst
duration is shorter than the cloud light-crossing time, which might be
the case for Sgr~C2 according to the rapid change observed in its
light curve (Fig.~\ref{fig:lightcurves}). Ultimately, the
line-of-sight geometry of the cloud would also be required in order to
translate the observed light curves into the original burst
profile. Unfortunately, no satisfying heuristic can be found to
estimate the influence of these parameters. It would require a refined
version of the Monte Carlo model, which is not yet available.

Despite these limitations, the position and column density we find for
Sgr~C2 agree well with those required to account for the extinction of
the north-eastern portion of the SNR candidate \snr
\citep{chuard2017}. It is also worth noting that while some of the
6.4~keV emission in Sgr~C4 is superposed upon the SNR candidate
emission, there is no sign of absorption of the SNR candidate emission
by Sgr~C4. As a consequence, \snr may consistently be located
somewhere between Sgr~C2 and Sgr~C4. Finally, our results also agree
with the work of \cite{sawada2004}, recently confirmed by
\cite{yan2017}, who derived a distribution of the molecular clouds in
the Galactic centre that is compatible with the positions we find for
the main subregions of Sgr~C.

Nevertheless, our results are significantly different from previous
estimates that were based on partial absorption of the local plasma
emission reported by \cite{ryu2013}. The origin of this discrepancy is
unclear. The model used by previous studies to describe the reflected
spectrum coming from the illuminated cloud (an absorbed power law) is
poorly suited, as was shown by \cite{walls2016}. In particular, the
strong dependence of the spectrum on the line-of-sight angle can
hamper the partial absorption measurement if it is not properly
modelled. Furthermore, we note that since the molecular cloud only
covers part of the region chosen for the spectral analysis, a fraction
of the flux is not in the line of sight of the cloud. Hence, part of
the absorption is independent of the cloud location, and it is
inaccurate to consider the cloud density liable for all the absorption
of the thermal plasma in the region. This is another important bias in
the derivation of the cloud position from the fitted absorption
fraction as done by \cite{ryu2013}.

\subsection{Investigation of the associated illuminating events}
\label{pastactivity}
Constraining the 3D positions of the clouds within the CMZ opens the
door to reconstructing the past light curve of \SgrA. To do so, we
developed a proxy of the Monte Carlo model to fit the time delay of
the associated echo directly instead of the line-of-sight
angle. Considering that Sgr~C1, C4 and B2 are roughly located on the
same parabola, we tested two hypotheses: $(i)$ that all clumps result
from the same event, and $(ii)$ that the illumination in Sgr~C2 is due
to a second event. Following the Occam razor parsimony principle, we
restricted our analysis to one-event and two-event models. We did not
consider models with more events to avoid overfitting.

We first fitted the spectra with the proxy model while imposing the
same value of the delay on all clumps. Cloud column densities were
fixed to the best-fit values we found when fitting each clump
individually to ensure stability in the parameter estimation. Then, we
repeated the procedure, now assuming that Sgr~C2 corresponds to a
different delay. We compared the statistics of these two fits using a
likelihood-ratio test and found that the second hypothesis is
statistically better than the first ($p<0.05$). The best-fit values of
the associated delays (2000 being the reference year) are $\Delta t_1
= 138_{-17}^{+27}$~yr for Sgr~C2 and $\Delta t_2 = 243_{-25}^{+20}$~yr
for all other clouds (Sgr~C1, C4, and B2). As the systematics
discussed before may affect these values, we repeated the same
analysis in the $Z=1.3$ case . We found again that the two-flare
scenario is preferred, and the delays are now $\Delta t_1 =
111_{-11}^{+14}$~yr and $\Delta t_2 = 204_{-16}^{+24}$~yr.

Another important assumption of the model is the cloud density
profile. However, as stated in Section~\ref{montecarlo}, alternative
Gaussian profiles are statistically less good, and in any case, they
cause the clouds to lie farther apart. This would be even more
inconsistent with a single-event scenario. Therefore our conclusions
are rather robust against systematic effects and the values obtained
in the $Z=1.3$ case provide an estimate of the systematic uncertainty
in the age determination of the outbursts.

The finding that Sgr~C2 is illuminated by a second event is further
supported by the two distinct variation patterns that can be
identified in the light curves derived in
Section~\ref{specanalysis}. The time behaviours of Sgr~C2 and Sgr~C4
are indeed very different, one exhibiting a sudden rise and fall in
about 2005, while the other decreases smoothly during the entire
period. Even assuming that the Fe\Ka emission in Sgr~C2 peaked in
about 2003, the illumination duration barely exceeds eight
years. Because of the delay due to the propagation, the timescale of
the associated flare has to be substantially shorter (i.e. a few
years). Conversely, the light curve of Sgr~C4 suggests a considerably
longer timescale (ten years at least). This difference strongly
supports the two-event scenario. Interestingly, \cite{terrier2017}
reported a trend for Sgr~B2 very similar to the trend of Sgr~C4, which
is an important hint that these two clumps may be witnessing the same
event. Although the light curve of Sgr~C1 shows no evidence of a
comparable linear decrease, the imaging analysis makes it clear that
it is also illuminated by a long outburst \citep[whose duration is
  estimated to $\sim20$~yr by][]{terrier2017}. As a consequence, the
evidence in support of the two-event scenario does not come from
position determination alone, but also from consistent patterns of
variation found in the images and the light curves.

\section{Discussion and conclusion}
\label{discussion}
Sgr~C is a much more complex region than Sgr~B2. It notably hosts a
SNR candidate that might be interacting with molecular material, as
well as two nearby pulsar wind nebula candidates. As a consequence,
this site is expected to be a good candidate for the cosmic-ray
irradiation scenario. However, we were able to provide significant
evidence of variability of the Fe\Ka emission in all
considered subregions. In this regard, it should be noted that Sgr~C2
is located very close to the sharp edge of \snr \citep{chuard2017},
which is a possible interaction region of the SNR candidate with
molecular gas, that is, a candidate site for intense cosmic-ray
production. Despite this, Sgr~C2 is the clump for which we
have the strongest evidence of short-term variability. This result,
along with the consistent interpretation given in
Section~\ref{montecarlo}, is a very strong argument in favour of the
reflection scenario. This does not exclude the possibility that LECR
production may still contribute to the nearly constant level of
background emission (see Fig.~\ref{fig:lightcurves}), however.

We propose that Sgr~C1, C4, and B2 were illuminated by the same event
that took place about 240 years before present\footnote{The time delay
  due to the propagation of light from the Galactic centre to Earth
  ($\sim 26\,000$~yr) is not considered here.} and lasted at least a
decade. Additionally, we find that Sgr~C2 was likely illuminated by a
second flare that took place about 100 years later and lasted no
longer than a few years. These results, which rely on imaging
analysis, light-curve extraction, and three-dimensional position
determination using a Monte Carlo spectral model, appear to be
consistent with previous works that used different approaches. The
most recent flare we report may be the same as the 110-year-old
outburst described by \cite{churazov2017}, based on the comparison of
time and space structure functions of the emission of a part of the
Sgr~A molecular complex. Furthermore, our findings are in very good
agreement with the two-event scenario proposed by \cite{clavel2013}
for the Sgr~A complex. As a result, the long flare seen in Sgr~C1, C4,
and B2 may be the same as the one seen by these authors in the
molecular clouds known as MC1, MC2, and G0.11$-$0.11, while the short
event seen in Sgr~C2 may be the same as the one they see in the
Bridge. Since we are able, for the first time, to assign an
approximate date to these outbursts, we can infer the possible
locations of the clumps in Sgr~A. If the Bridge were indeed
illuminated by the same flare as Sgr~C2, the $z$-coordinates (as
defined in Fig.~\ref{fig:geom}) of its subregions Br1 and Br2 would
thus be in the range of $z\sim 5 - 25$~pc. Similarly, assuming that
MC1, MC2, and G0.11$-$0.11 recently witnessed the 240-year-old flare,
their $z$-coordinates would thus be in the range of $z\sim 20 -
45$~pc. Following this approach, these predictions can be considered a
test of the two-event scenario.

We are able to provide only lower limits on the luminosity of the
associated outbursts. The implied luminosity varies with the inverse
square of the radius of the cloud, assumed spherical, which is very
poorly constrained. We used the size of the spectral extraction region
instead, meaning that we clearly underestimate the
luminosity. Moreover, clouds may not be fully illuminated, contrary to
what we assume. It is also probable that we do not have data that
match the time of maximum emission. Bearing this in mind, the X-ray
luminosity (2--10~keV) of the two events illuminating Sgr~C is
estimated to be at least a few $10^{38}$~\ergs, and perhaps
significantly higher (up to a few $10^{39}$~\ergs). As was extensively
discussed in \cite{clavel2013} for the Sgr~A complex, \SgrA is the
best candidate to account for the luminosity, spectral index, and
flare duration required to explain the Fe\Ka emission observed in
Sgr~C and to consistently illuminate both sides of the CMZ in three
independent molecular complexes. The two flares we report could be due
to stochastic variations of the accretion rate \citep{cuadra2008} or
tidal disruption events \citep[TDE; see e.g.][]{zubovas2012}. While
the constraints we place on their age and frequency may help in
investigating their origin, both further regular monitoring of the
Fe\Ka emission of the CMZ and greater modelling efforts are still
needed to ultimately unveil the past light curve of \SgrA.


\begin{acknowledgements}
This research has made use of data obtained with \xmm, an ESA science
mission with instruments and contributions directly funded by ESA
Member States and NASA, and from the \chandra Data Archive, as well as
software provided by the \chandra X-ray Center (CXC) in the
application packages CIAO, ChIPS, and Sherpa. The authors acknowledge
the Centre National d'\'Etudes Spatiales (CNES) for financial
support. GP acknowledges support from the Deutsches Zentrum für Luft-
und Raumfahrt (DLR).

\end{acknowledgements}

\bibliographystyle{aa} 
\bibliography{sgrC.bib} 

\end{document}